# IMAGE ENCRYPTION BASED ON DIFFUSION AND MULTIPLE CHAOTIC MAPS

[1]G.A.Sathishkumar ,[2]Dr.K.Bhoopathy bagan  and [3]Dr.N.Sriraam

[1] Associate Professor, Department of Electronics and Communication Engineering,
Sri Venkateswara College of Engineering,Sriperumbudur -602108.
`sathish@svce.ac.in`

[2]Professor and HEAD, Department of Instrumentation , Madras Institute of Technology,
Chrompet, Chennai-600044
`Kbb02@yahoomail.com`

[3] Professor and Head, Department of Biomedical Engineering
SSN College of Engineering, Chennai 603110

## ABSTRACT

*In the recent world, security is a prime important issue, and encryption is one of the best alternative way to ensure security. More over, there are many image encryption schemes have been proposed, each one of them has its own strength and weakness. This paper presents a new algorithm for the image encryption/decryption scheme. This paper is devoted to provide a secured image encryption technique using multiple chaotic based circular mapping. In this paper, first, a pair of sub keys is given by using chaotic logistic maps. Second, the image is encrypted using logistic map sub key and in its transformation leads to diffusion process.  Third, sub keys are generated by four different chaotic maps. Based on the initial conditions, each map may produce various random numbers from various orbits of the maps. Among those random numbers, a particular number and from a particular orbit are selected as a key for the encryption algorithm. Based on the key, a binary sequence is generated to control the encryption algorithm. The input image of 2-D is transformed into a 1- D array by using two different scanning pattern  (raster and Zigzag ) and then divided into various sub blocks. Then the position permutation and value permutation is applied to each binary matrix based on multiple chaos maps. Finally the receiver uses the same sub keys to decrypt the encrypted images. The salient features of the proposed image encryption method are loss-less, good peak signal –to noise ratio (PSNR), Symmetric key encryption, less cross correlation, very large number of secret keys, and key-dependent pixel value replacement.*

## 1. INTRODUCTION

In recent years, more and more consumer electronic services and devices, such as mobile phones and PDA (personal digital assistant), have also started to provide additional functions of saving and exchanging multimedia messages [10], [11], [13]. The prevalence of multimedia technology in our society has promoted digital images and videos to play a more significant role than the traditional dull texts, which demands a serious protection of users' privacy. To fulfil such security and privacy needs in various applications, encryption of images and videos is very important to frustrate malicious attacks from unauthorized parties. Due to the tight relationship between chaos theory[14],[15] and cryptography, chaotic cryptography have been extended to design image and video encryption schemes

## 1.1 the Need for Image Encryption Schemes

The simplest way to encrypt an image or a video is perhaps to consider the 2-D and 3-D stream as a 1-D data stream, and then encrypt this 1-D stream with any available key , such a simple idea of encryption is called naive encryption[7],[20]. Although naive encryption is sufficient to protect digital images and videos in some civil applications, this issues have taken into consideration when advanced encryption algorithms are specially designed for sensitive digital images and videos, for their special features are very different from texts.

The recent research activities in the field of nonlinear dynamics and especially on systems with complex (chaotic) behaviours [3], [14] have forced many investigations on possible applications of such systems. Today, chaotic encryption [4],[5],[6],[7] is almost exclusively considered inside the nonlinear systems community.

## 1.2. Chaotic Maps

Chaos theory [3], [14], [15] describes the behaviour of certain nonlinear dynamic system that under specific conditions exhibit dynamics that are sensitive to initial conditions. The two basic properties of chaotic systems are the sensitivity to initial conditions and Mixing Property. In this paper, 1 D [15] chaotic map is used to produce the chaotic sequence and used to control the encryption process. The chaos streams are generated by using various chaotic maps. Among the various maps, four maps are investigated and their characteristics are analyzed.

### 1.2.1. Logistic Map

A simple and well-studied example [3], [14] of a 1D map that exhibits complicated behavior is the logistic map from the interval $[0,1]$ in to $[0,1]$, parameterised by µ:

$$g_\mu(x) = \mu * (x) \qquad (1)$$

The state evolution is described by $x(n+1)=\mu*x(n)*(1-x(n))$ (2)

Where $0 \leq \mu \leq 4$. This map constitutes a discrete-time dynamical system in the sense that the map $g_\mu : [0,1] \to [0,1]$ generates a semi-group through the operation of composition of functions. In the logistic map, as µ is varied from 0 to 4, a period-doubling bifurcation occurs.

### 1.2.2. Tent Map

In mathematics, the tent map [3],[14] is an iterated function, in the shape of a tent, forming a discrete-time dynamical system. It takes a point $x_n$ on the real line and maps it to another point:

$$x_{n+1} = \begin{cases} \mu x_n & \text{for } x_n < \tfrac{1}{2} \\ \mu(1 - x_n) & \text{for } \tfrac{1}{2} \leq x_n. \end{cases} \qquad (3)$$

Where µ is a positive real constant

Depending on the value of µ, the tent map demonstrates a range of dynamical behavior ranging from predictable to chaotic.

### 1.2.3. Quadratic Map

More complicated analytic quadratic map [3],[14] is

$$x_{n+1} = f_c(x_n) = x_n^2 + c \qquad (4)$$

For an analytic map points where f '($x_c$) = 0 are called critical points. Quadratic map has the only critical point $x_c$ = 0.So a fixed point is stable (attracting), super stable, repelling, indifferent (neutral) according as its multiplier satisfies |m| < 1, |m| = 0, |m| > 1 or |m| = 1. The second fixed point is always repelling. For |x| > $x_2$ iterations go to infinity. For |x| < x2 they go to the attracting fixed point x1. This interval is the basis of attraction of the point.

**1.2.4. Bernoulli Map**

Bernoulli map [3],[14] or the 2x mod 1 map defined as

$$f(x) = \begin{cases} 2x, & 0 \leq x < 0.5 \\ 2x-1, & 0.5 \leq x < 1 \end{cases} \qquad (5)$$

A Bernoulli process is a discrete time stochastic process consisting of a finite or infinite sequence of independent random variable $X_1, X_2, X_3,...$, such that for each i, the value of $X_i$ is either 0 or 1; for all values of i, the probability that $X_i = 1$ is the same number p. From any given time, future trials are also a Bernoulli process independent of the past trails.

## 2. MUTIMAP ORBIT HOPPING

In this paper, the chaotic key generation is shown in block diagram [19] Figure 1. Given a key, the hopping mechanism performs a key handling process, then chooses *m* maps *M0, M1 …, Mm-1* from the chaotic map bank and sets the order of the chosen maps to hop. In addition, for each individual chosen map, *s* orbits S0, S1, …, Ss-1 are generated. Further, on each orbit, *n* points N0, N1, …, Nn-1 are generated. The key determines parameters m, s, n, and hopping pattern. For a given chaotic map, the second orbit is generated by increasing the initial seed of the first orbit by an offset.

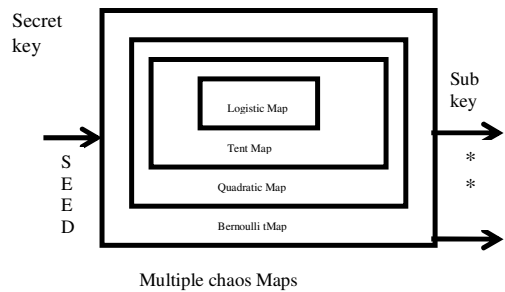

**Figure 1.** Typical architecture of Multi-Map Orbit Hopping Chaotic Key Generation

The third orbit is generated by increasing the initial seed of the second orbit by the same offset, and so on so forth for other extra orbits needed. In this paper, 1 D chaotic map is used to produce the chaotic sequence and used to control the encryption process. The maps used in this paper [3],[14],[15] are logistic map, Bernoulli map, Tent map and Quadratic map for key generation. The chaos streams are generated by using various chaotic maps.

## 3. THE PROPOSED IMAGE SECURITY SYSTEM

The proposed encryption algorithm belongs to the category of the combination of value transformation and position permutation [7]. In this paper, two different types of scanning methods are used and their performances are analyzed. . The typical schematic of the proposed method is shown in Figure 2. First, a pair of sub keys is given by using chaotic logistic maps. Second, the image is encrypted using logistic map sub key and in its transformation leads to diffusion process.

Third, sub keys are generated by four different chaotic maps and images are treated as a 1D array by performing Raster scanning and Zigzag scanning. The scanned arrays are divided into various sub blocks. Then for each sub block, position permutation and value transformation are performed to produce the encrypted image. The sub keys are generated by applying the suitable chaotic map banks. Based on the initial conditions, the generated chaotic map banks are allowed to hop through various orbits of chaotic maps. The hopping pattern [19] is determined from the output of the previous map. Hence, for each sub block various chaotic mapping patterns are applied which further increases the efficiency of the key to be determined by the brute force attack. In each orbit, a sample point is taken and used as key for a specific block and a condition to choose the particular orbit in a particular map is adopted. Then, based on the chaotic system, binary sequence is generated to control the bit-circulation functions for performing the successive data transformation on the input data.

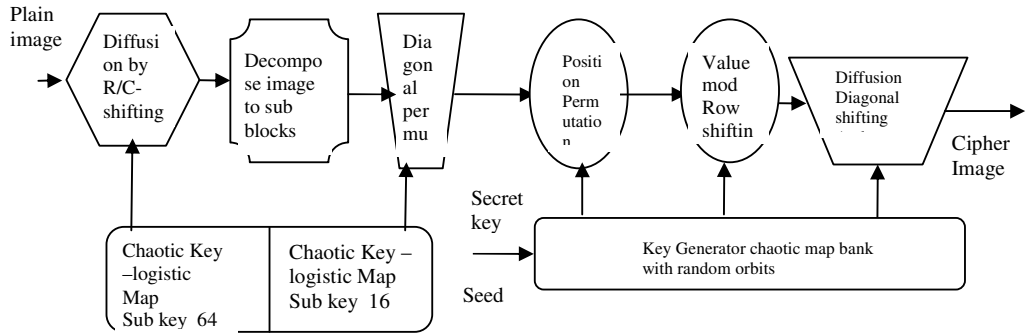

**Figure2.** Proposed Chaos based image cryptosystem

Eight 8-bit data elements are regarded as a set and fed into an 8 × 8 binary matrix. In the successive transformation on each diagonal by using these two functions, we randomly determine the two parameters used in the functions according to the generated chaotic binary sequence such that the signal could be transformed into completely disorderly data. In addition to chaotic features of mixing, unpredictable, and extreme sensitive to initial seeds, through chaotic maps and orbits hopping mechanism, we spread out the pseudo random number base to a wide flat spread spectrum in terms of time and space. The following steps carried out for the implementation [5][7][19] of proposed chaos based mapping technique.

Let' s denote a one-dimensional (1D) [5]digital signal of length N, s(n), $0 \leq n \leq N - 1$, be the one-byte value of the signal s at n, M an 8×8 binary matrix, and s' and M' the encryption results of s and M, respectively. In the following definitions, the integer parameters r and s are assumed larger than or equal to 0, but they are less than 8.

### 3.1 Algorithm-I – Diffusion Process

Step 1: Break down the image into 8 x 8 sub blocks. (a total of 64 x 64 sub blocks will emerge).
Step 2: Perform Row wise and Column wise rotation based on the chaos key as follows ( see Figure 3)

**Row Wise Rotation:**
If the bit in the chaos key is '1' then perform row rotation (one rotary shift to the right). If the bit in the chaos key is '0' then perform column rotation (one rotary shift of the column downward) on the Column. Each key can be used for each row. The next key will be used on the next row and so on. But in with all the keys only the first column is rotated when the key bit is '0'. A total of 64 keys (one for each row) will be required for this Row Wise Rotation

**Column Wise Rotation:**
If the bit in the chaos key is '1' then perform row rotation (one rotary shift to the right) on the Row number 1 .If the bit in the chaos key is '0' then perform column rotation (one rotary shift of the column downward). Each key can be used for each column.
The next key will be used on the next column and so on. A total of 64 keys (one for each column) will be required for this Column Wise Rotation
Step 3: Following the Row wise and Column wise rotation the newly generated sub block arrangement is used to recompose the image. The Recomposed image is not Block permuted based on row wise and column wise rotation
Step 4: the Block Permuted image is now decomposed into 64 x 64 image sub blocks.

Step 5: Each 64 x 64 sub block is then subjected to diagonal rotation. A total of 16 keys will be required for diagonal rotation (one for each sub block)

Step 6: If the chaos key bit in the key corresponding to a specific sub block is '0', then each diagonal in from top left to bottom right in the sub block is rotated by one position

If the chaos key bit in the key corresponding to a specific sub block is '1', then each diagonal in from top right to bottom left in the sub block is rotated by one position. This step performs the necessary pixel permutation

Step 7: from the pixel permuted 64 x 64 sub blocks the image is then recomposed. This gives the encrypted image

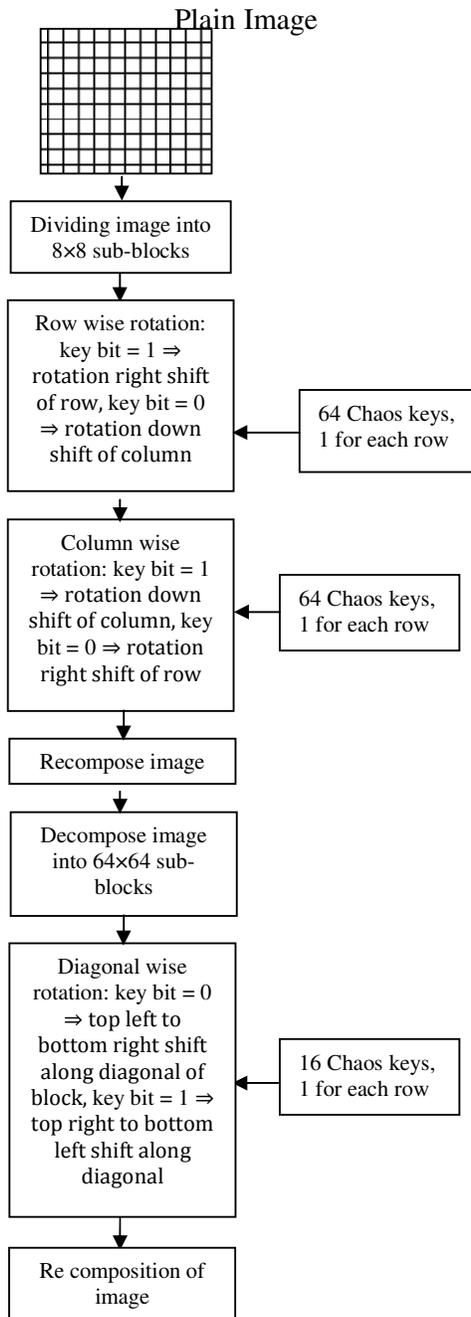

**Figure 3. Diffusion Process**

Decryption:
Decryption operation is similar to the encryption operation. The only differences being that the key is traversed in the reverse direction rather than the forward direction and the rotations based on the key bits are performed in a direction opposite to that used in Encryption. For Eg. if the in encryption the row was rotated right-ward, then in decryption it is rotated left-ward. And in order to retain the correct sequence of rotation, the key is traversed in the reverse direction in all the rotation loops.

### 3.2 Algorithm-II- Multiple Maps

Step1: Covert 2-D image into 1-D array then perform a) Raster scanning and b) Zigzag scanning. ( see Figure 4)
Step2: Consider a block size of 8 x 8 and convert them in to binary values.
Step3: Sub key size are at least 20 bits, it is extracted from the chaos map banks. The Secret key is SEED, which are the initial conditions of the each map. Based on the initial conditions the chaotic map banks are allowed to hop through various orbits of different chaotic maps. The hopping pattern is determined from the output of the previous map. Then, based on the chaotic system, binary sequence generated to control the bit-circulation functions for performing the successive data transformation on the input data. Given pair of $f$ and $f'$, the combination of $p, q, r, t, u$ and $s$ resulting in the transformation pair may be non-unique which is the secret key.

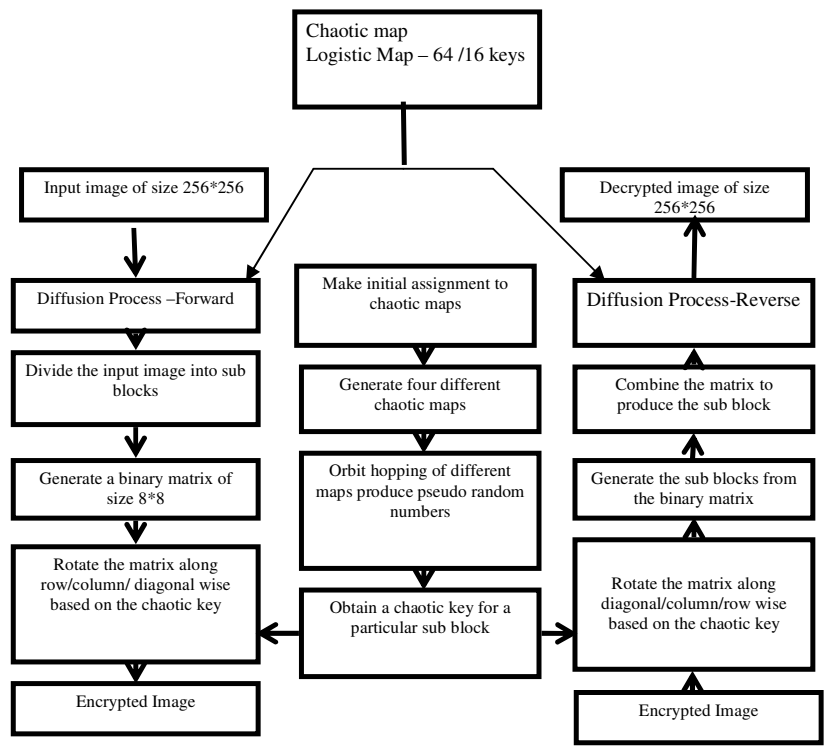

**Figure 4.** Typical architecture of the proposed chaos based image crypto systems

Step4: Convert the chaotic sub key in to binary values of 20 bits.
Step5: Each 8x 8-sub block of image pixel values circularly shifted with chaos map banks.

Step6: **Definition for 2 D Circular Shifting of Diagonal pixels (CSDP):** The Mapping [5] $ROLR_k^{t,u}$ & $ROD_k^{t,u} f \to f'$ is defined to rotate each pixel at the position (x ,y) in the image such that $k^{th}$ diagonal of $f$ $0 \leq f \leq 7$, u bits in the up direction if t equals 1 or u bits in the down direction if t equals 0. In different combinations of p, q, r, t, u and s, the composite mapping is given below

$$(\sum_{j=0}^{7} ROLR_j^{q,s}) (\sum_{i=0}^{7} ROLR_i^{p,r}) \cdot (\sum_{k=0}^{13} ROLR_k^{t,u}) \qquad - (6)$$

The proposed method (**CSDP**) possesses the following desirable features:

A binary matrix $f$ be transformed into quite different matrixes and different matrixes can be transformed into the same matrix. Given a transformation of pair $f$ and $f'$ the combinations of p, q, r and s resulting in the transformation pair may be non-unique.

Since $f$ is an 8 × 8 matrix, the result of circulating diagonal k bits and of circulating it (kmod8) bits in the same direction. This is why r and s are assumed to be in the ranges of $0 \leq r \leq 7$ and $0 \leq s \leq 7$.

Step 7: Perform the encryption based on the chaotic key values, which is obtained from the different orbits of chaos maps chosen by hopping randomly.
Step 8: Transform the encrypted image 1-D to 2-D.
Step 9: Transmit the Chaotic sub key via secure channel using public key algorithms.
Step10: Decrypt the cipher image using the same chaotic sub key and SEED.
Finally, carry out performance analysis by doing correlation, histogram, speed and loss of the original, encrypted and decrypted image.

### 3.2 ANALYSIS OF SECURITY PROBLEM FOR A MUTLI – MAP HOPPING

For an unknown set of μ and x(0) of the logistic map, the number of possible encryption results is $2^{16}$N/8 if the TDCEA [5] is applied to a signal of length N**.** Since it requires 16N/8 bits to encrypt a signal of length N, the number of possible encryption results is $2^{16}$N/8. Since the chaotic binary sequence is unpredictable and furthermore, proposed technique (**CSDP**) multi – hopping chaotic sequence are used therefore, it is very difficult to decrypt correctly an encrypted signal by making an exhaustive search without knowing μ and x(0). Moreover, small fluctuation in μ and x(0) results in quite different chaotic binary sequence because the trajectory of the chaotic system is very sensitive to initial condition.

By the way of collecting, some original signals and their encryption results or collecting some specified signals and their corresponding encryption results, it is impossible for the crypt analyst to decrypt correctly an encrypted image without knowing μ and x(0). Because the rotation direction and the shifted bit-number in each row or column transformation randomly determined by the multi - hopping chaotic binary sequence. Hence, the new scheme (**CSDP**) can resist the chosen cipher text attack and the known plaintext attack.

## 4. Experimental Results

An image size of 256 * 256 (e.g. cameraman, pepper, aero, etc.,) is considered as plain (original) image and CSDP is performed with multi map orbit key. The most direct method to decide the disorderly degree of the encrypted image is by the sense of sight. On the other hand, the correlation coefficient can provide the quantitative measure on the randomness of the encrypted images.

In order to apply the (CSDP)**,** the parameters α and β must be determined according to Step 1. Based on the experimental experience, general combinations of α and β can always result in very disorderly results. In the simulation, α = 2 and β = 2 are adopted in Step 1. The initial conditions of all chaotic maps used are set as, x(0) = 0.75 and μ = 3.9 for logistic map, c=1.75 for tent map, f(x)=0.5 for Bernoulli map and finally a=.5 ,b=.25 for quadratic map. The offset values for producing various orbits are chosen to be very less than the initial conditions.

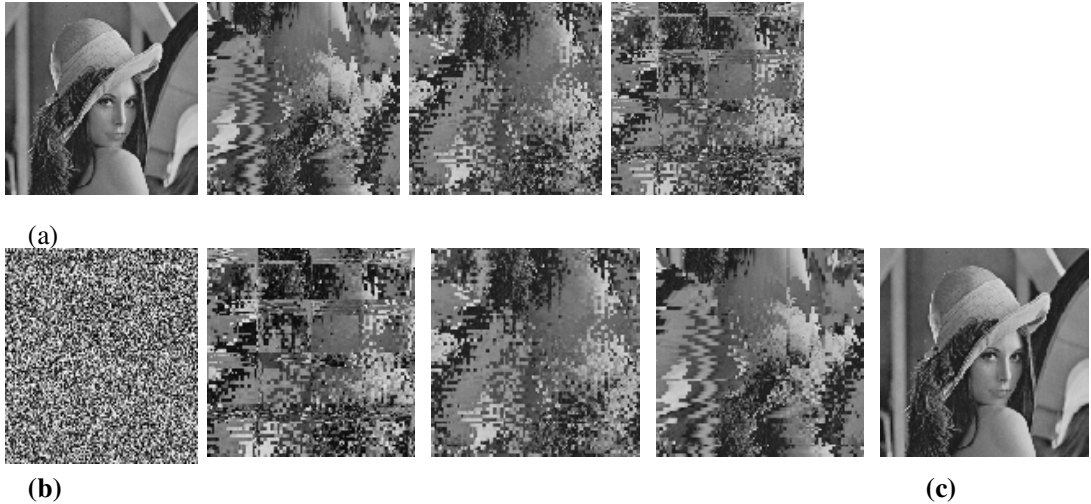

(a)

(b) (c)

**Figure 5.** (a) Plain image   (b) Encrypted image   (c) Decrypted image

The visual inspection of Figure 5, shows the possibility of applying the algorithm successfully in both encryption and decryption. In addition, it reveals its effectiveness in hiding the information contained in it.

### 4.1 Histogram Analysis

To prevent the leakage of information to an opponent [2],[4],[6] it is also advantageous if the cipher image bears little or no statistical similarity to the plain image.

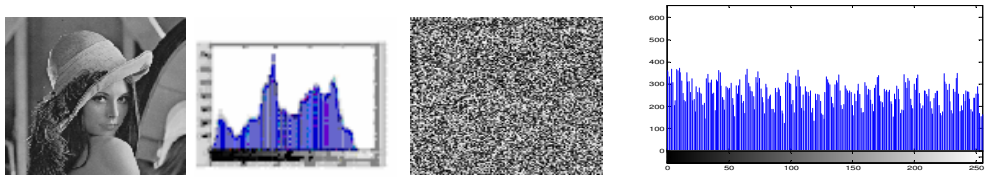

**Figure 6.** Histogram of original and encrypted image

We have calculated and analyzed the histograms of the several encrypted images as well as its original images that have widely different content. One typical example among them is shown in Figure 6. The histogram of a plain image contains large spikes. The histogram of the cipher image is shown in Figure 6, it is uniform, significantly different from that of the original image, and bears no statistical resemblance to the plain image and hence does not provide any clue to employ any statistical attack on the proposed image encryption procedure

### 4.2 Correlation Co Efficient Analysis

In addition to the histogram analysis [4],[6], we have also analyzed the correlation between two vertically adjacent pixels, two horizontally adjacent pixels and two diagonally adjacent pixels in plain image and cipher image respectively. The procedure is as follows:

$$r_{xy} = \frac{cov(x,y)}{\sqrt{D(x)}\sqrt{D(y)}} \qquad (7)$$

Where x and y are the values of two adjacent pixels in the image. Figure 7, shows the correlation distribution of two horizontally adjacent pixels in plain image and cipher image.
The correlation coefficients are 0.9905 and 0.0308 respectively for both plain image and cipher image. Similar results for diagonal and vertical directions. It is clear from the Figure.4.3 and Table 1 that there is negligible correlation between the two adjacent pixels in the cipher image. However, the two adjacent pixels in the plain image are highly correlated. The correlation coefficients of various maps are calculated and they are compared with each other. The comparison table for various plain images, various cipher images and various maps based on the correlation coefficient are calculated and tabulated in Table 1 and 2.

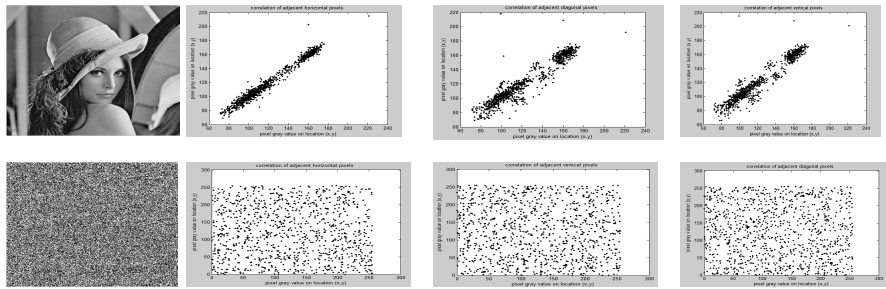

**Figure 7.** Horizontal, vertical and diagonal correlation of plain and cipher image

**Table 1** Horizontal, Vertical & Diagonal Correlation of Encrypted Image

| plain Image | Algorithm II(CSDP) | | | Algorithm I and II(Diffusion and CSDP) | | |
|---|---|---|---|---|---|---|
| | Correlation(H) | Correlation (V) | Correlation (D) | Correlation (H) | Correlation (V) | Correlation (D) |
| Elaine | -0.0139 | -0.0954 | -0.0710 | 0.0237 | -0.0432 | 0.0090 |
| Barbara | 0.0286 | 0.0230 | -0.0244 | -0.0031 | 0.0324 | -0.0169 |
| Pepper | -0.0025 | -0.0219 | 0.0167 | 0.0093 | -0.0308 | -0.0381 |
| Lena | -0.0590 | -0.0381 | -0.0457 | -0.0332 | 0.0608 | 0.0567 |
| Cameraman | 0.0286 | 0.0230 | -0.0244 | -0.0432 | 0.0305 | 0.0275 |
| Airfield | 0.0075 | 0.0069 | -0.0317 | 0.0802 | -0.0300 | -0.0095 |

**Table 2** Correlation coefficient for various different maps after zigzag scanning (Diffusion and CSDP)

| Images | Logistic Map | Bernoulli Map | Tent Map | Quadratic Map |
|---|---|---|---|---|
| Elaine | -0.0011 | -0.00030107 | -0.000669 | -0.00033906 |
| Lena | -0.0107 | -0.0110 | -0.0028 | -0.0110 |
| Bridge | -0.0065 | -0.0026 | -0.0041 | -0.0027 |
| Airfield | 0.00038084 | 0.0014 | 0.00023981 | 0.0015 |
| Peppers | -0.002 | **-0.000087834** | 0.00054437 | -0.00012799 |
| Cameraman | 0.0034 | -0.0049 | 0.0019 | -0.0049 |
| Barbara | -0.0079 | -0.0089 | -0.0032 | -0.0089 |

**Table 3** Cross Correlation coefficients for raster scanning and zigzag scanning

| Plain Image | Algorithm II(CSDP) | | Algorithm I and II(Diffusion and CSDP) | |
| --- | --- | --- | --- | --- |
| | Raster Scanning | Zigzag Scanning | Raster Scanning | Zigzag Scanning |
| Elaine | 0.0539 | -0.0139 | -0.0089 | -0.0048 |
| Lena | -0.0535 | -0.0590 | 0.0074 | -0.0080 |
| Bridge | 0.0174 | -0.0023 | -0.0111 | -0.0004938 |
| Airfield | -0.0213 | 0.0075 | -0.0036 | -0.0034 |
| Barbara | 0.0901 | 0.0286 | -0.0195 | -0.0753 |
| Peppers | 0.0901 | -0.0025 | 0.0057 | 0.00027851 |

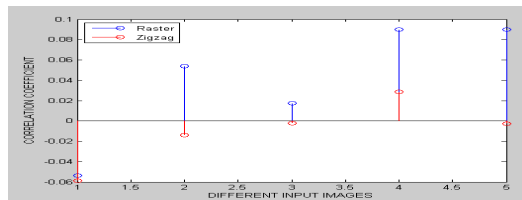

**Figure 8.** Cross correlation for different Scanning Patterns(with CSDP)

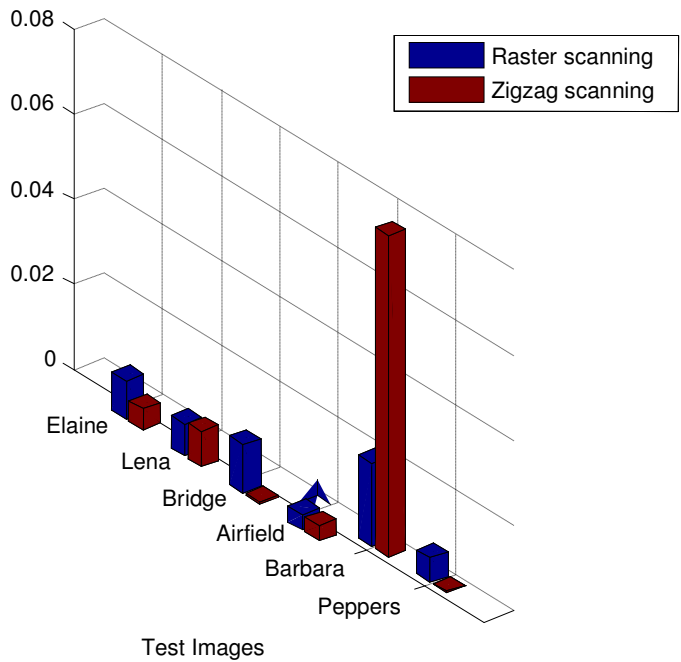

Figure 9. Absolute value of cross correlation (with CSDP & Diffusion process)

**Table 4** Correlation Coefficients in Plain image and Cipher image

| Direction of Adjacent Pixels | Plain image | Cipher image using single map | Cipher image using multiple map | Cipher image with CSDP & Diffusion |
|---|---|---|---|---|
| Horizontal | 0.9670 | 0.0781 | -0.0025 | -0.0031 |
| Vertical | 0.9870 | 0.0785 | -0.0218 | 0.0324 |
| Diagonal | 0.9692 | 0.0683 | 0.0167 | -0.0160 |

The correlation coefficients are found for the various directions of scanning patterns employed and the tabulated in the Table 3.In Table 4, the observation shows that the zigzag scanning is more efficient than the raster scanning (See Figure 8 and 9). In addition, cipher image with multiple maps are more resistant to cryptanalytic attacks.

**Table 5** The encryption speed of the proposed .The analysis has been done on an Intel Core 2 Duo 2.1 GHz CPU with 2 GB RAM running on Windows Vista Home Basic and using the MATLAB 7.2 a programming.

| Image | Sub block Size | Speed (sec's) | Algorithm –II | Algorithm I and II (Diffusion and CSDP) | | |
|---|---|---|---|---|---|---|
| | | | PSNR dB | Entropy | PSNR dB | Cross correlation |
| Bridge | 8x8 | 108.854503 | 9.2637 | 7.9413 | 8.47 | -0.0004938 |
| Lena | 8x8 | 106.908351 | 9.2006 | 7.9634 | 8.97 | -0.0080 |
| cameraman | 8x8 | 111.030547 | 8.3399 | 7.8472 | 8.34 | -0.0046 |

## 4.3 Key sensitivity analysis

The key sensitivity is an essential feature for any good cryptographic algorithm which guarantees the security of the proposed system against the brute-force attack to great extent. The key sensitivity of a proposed method can be observed in two different ways: (i) the cipher image should be very sensitive to the secret key, i.e., if we use two different keys with single bit deviation to encrypt the same plain image then two cipher images produced should be completely independent to each other and they should possess negligible correlation and (ii) the cipher image cannot be decrypted correctly if there is a small difference between the encryption and decryption keys. As there are four different maps are involved (four floating point numbers and one integer as seed ) in the secret key of the proposed encryption/decryption technique and diffusion process needs a (single floating point number ), so we have tested the sensitivity with respect to each part of the secret key and the proposed algorithm gives less correlation( Table 5).

## 4.4 Plaintext sensitivity analysis (differential analysis)

In order to test known plain text attack and chosen plaintext attack, a cryptanalyst attempts to make a one bit change ,i.e., usually one pixel, in the plain image and compare the cipher images to extract useful relationship between plain image and cipher image, which further determines the secret key. Such analysis is called as differential cryptanalysis [4][8][22] in cryptography. If a small change in the plain image causes significant changes in the cipher image then such differential analysis may become inefficient. To test the influence of one pixel change on the whole cipher image, two most common measures NPCR (net pixel change rate) and UACI (unified average changing intensity) are used.

Let two ciphered images, whose corresponding plain images have only one pixel difference; be denoted by CI1 and CI2.Label the grayscale values of the pixels at grid (i,j) in CI1 and CI2 by C I(i,j) and CI(i,j), respectively. Define a bipolar array D, with the same size as images CI1 and CI2. Then, Diff(i,j)is determined by CI1(i,j) and CI2(i,j), namely, if CI1(i,j) =CI2(i,j) then Diff(i,j) = 1; otherwise, Diff(i,j) = 0.The NPCR [21][22] is defined as

$$NPCR = \frac{\sum_{i,j} Diff(i,j)}{W \times H} \times 100\% \qquad (8)$$

Unified average changing intensity (UACI) means changing intensity of the corresponding pixels of the plain image and cipher image. The larger the UACI is, the more resistant to the differential attack the encryption
scheme. The UACI [21] [22] is defined by:

$$UCAI = \frac{1}{W \times H} \left[ \sum_{i,j} \frac{CI1(i,j) - CI2(i,j)}{255} \right] \times 100\% \qquad (9)$$

The NPCR value for two random images, which is an expected estimate for a good encryption technique, is given by 99.69 and UACI 33.4653. For a proposed method NPCR value and UCAI are found to be 98.4754 and 32.2128.The proposed image encryption technique shows extreme sensitivity on the plaintext and hence it is not vulnerable to the differential attacks.

### 4.5 PSNR

PSNR [21][22] of encrypted image and original image is computed as follows

$$PSNR = 10 \log_{10} \frac{hw \left[ \max_{1<i\leq m, 1<j\leq n} \{p'_{i,j}\} \right]^2}{\sum_{i=1}^{h} \sum_{j=1}^{w} (p_{i,j} - p'_{i,j})^2} \qquad (10)$$

Where h and w are the width and height of original image, while pij and p'ij are pixel values of encrypted image and original image respectively.In the proposed scheme, higher the visual quality of the cipher image is, the less the number of changed pixels will be, and the larger the value of PSNR will be, it is around 9.2 for the CSDP (Algorithm –I) and 8.3 for the Combined CSDP and diffusion (Algorithm –I and II).

## 5. CONCLUSION AND FUTURE SCOPE

In addition to chaotic features of mixing, unpredictable, and extreme sensitive to initial seeds, through multiple chaotic maps and orbits hopping mechanism, we spread out the pseudo random number base to a wide flat spread spectrum in terms of time and space. It is similar to say that our pseudo random numbers are out of the white noise. Chaotic maps are computationally economic and fast. This proposed chaos based image cipher will be suitable for applications like wireless communications. In future, the proposed crypto system will be implemented and tested in the FPGA hardware.

## Authors

G.A.Sathishkumar Obtained his M.E degree from PSG College of Technology , Coimbatore ,India .He is currently pursuing PhD in Anna University,Chennai and Faculty member in the Department of Electronics and Communication Engineering ,Sri Venkateswara College of Engineering ,Sriperumbudur.His research interest is Network Security ,Image Processing ,VLSI & Signal Processing Algorithms

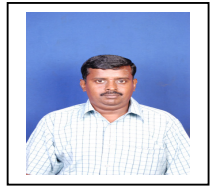

Dr.K.Bhoopathy Bagan completed his doctoral degree from IIT Madras. He is presently working as professor, ECE dept, in Anna University, MIT Chrompet campus, Chennai. His areas of interest include Image Processing, VLSI, Signal processing and network Security.

Dr.N.Sriraam He is presently working as Professor and Head Department of Biomedical Engineering SSN College of Engineering, Chennai 603110.